\documentclass[a4paper,11pt]{article}
\usepackage{amsmath,amssymb}
\usepackage{cite}

\newcommand{\email}[1]{\footnote{email:#1}}

\newcommand{\be}{\begin{equation}}
\newcommand{\ee}{\end{equation}}

\begin{document}

\title{Stability of the Scalar Mass against Loop Corrections}
\author{Kang-Sin Choi\email{kangsin@ewha.ac.kr}
 \\ \it \normalsize 
Scranton Honors Program, Ewha Womans University, Seoul 03760, Korea \\
\it \normalsize  
Institute of Mathematical Sciences, Ewha Womans University, Seoul 03760, Korea}
\date{}

\maketitle

\begin{abstract}
We show that the renormalized loop corrections to the scalar mass are suppressed if the field in the loop is heavy. Here, treating the renormalized mass is important, as it is the only observable. This means that the physics of a low-energy scalar field is insensitive to that of ultraviolet.
\end{abstract}

\vskip 1cm

The quantum theory of a scalar field exhibits the hierarchy problem in the mass. Although the mass, e.g., of Higgs, that we observe in the experiment is small compared to the ultraviolet (UV) scale, it has not been theoretically well understood. A bare mass $m_B^2$ that is written in the Lagrangian is modified by loop corrections due to interactions
\be \label{rendmass}
 m_B^2 + \sum_n \tilde \Sigma_n(p^2),
\ee
where $\tilde \Sigma_n(p^2)$ is the so-called self-energy whose Feynman diagrams are one-particle-irreducible and contain two external scalar legs, and $p^2$ is the size of the external momentum. We collectively denoted the dependence of various couplings by their orders $n$. For instance, a theory with a single coupling $g$ gives $\tilde \Sigma_n(p^2) \propto g^n$. It does not seem stable against such corrections because the corrections should scale like the energy squared characterizing the UV physics that is much larger than the observed mass. 

Renormalization renders a physical quantity finite; after all, there is no dependence on regularized infinity like loop momentum cutoff $\Lambda^2$ \cite{Zimmermann:1969jj, Choi:2024cbs} (see (\ref{MassCorr}) below). Yet, it does not explain the smallness of the mass. It is the mass $M$ of a heavy field in the loop that really characterizes the UV scale. The technical hierarchy problem has been formulated as follows based on dimensional analysis (see (\ref{selfeform}) below).
\begin{quote}
Old formulation: The quantum correction at each order 
\be \label{quaddiv}
\tilde \Sigma_n(p^2) = {\cal O}( M^2 \log M^2)
\ee
depends quadratically on $M^2$ and thus is sensitive to unknown parameters in the UV. Even if we force the mass to be finite and hierarchically smaller than $M^2$ at a certain order, higher-order corrections in general ruin it \cite{Gildener:1976ai,Gildener:1979dd,Georgi:1993mps,Georgi:1981vf,Arkani-Hamed:1998jmv,Weinberg:1978ym,Giudice:2013yca, NC}.
\end{quote}
This is of its own interest, whose best guiding principle has been the 't Hooft naturalness \cite{tHooft:1979rat}. We are interested in the {\em observed mass.} 

In the experiment, {\em only the combination (\ref{rendmass}) is observable}, while the unrenormalized correction (\ref{quaddiv}) is not separately observable \cite{Choi:2023cqs}.
What we do by renormalization is to change the mass basis
\be \label{SlidingMass}
%\begin{split}
 m^2(p^2) = m^2(\mu^2) + \sum_n \left[ \tilde \Sigma_n(p^2) -\tilde \Sigma_n(\mu^2) - (p^2-\mu^2)\frac{d\tilde \Sigma_n}{d (p^2)}(\mu^2) \right] 
 %\\
%\end{split}
\ee
that is, to separate the bare mass and the kinetic terms into the counterterms
\be
 m_B^2 \equiv m^2 (\mu^2) - \sum_n \tilde \Sigma_n(\mu^2).
\ee
(The last terms in (\ref{SlidingMass}) come from the field-strength renormalization.)
Note that this represents a reparametrization of the mass basis, which can equivalently define a one-parameter (for the mass) scalar theory.  Like gauge couplings, quantum field theory and renormalization teach us that the mass $m^2(p^2)$ is scale-dependent. For convenience, we deal with the pole mass $m_{\rm p}^2$ defined by the renormalization condition $m_{\rm p}^2 = m^2(m_{\rm p}^2)$ and drop the subscript. 
Note that we guarantee that the term in (\ref{SlidingMass}) at each order is finite after renormalization since the asymptotics in the high momentum regime do not depend on $p^2,\mu^2$. The only problem is that each term in (\ref{SlidingMass}) still depends on the UV mass $M^2$.

The above discussion leads us to reformulate the problem about the whole mass $m^2(p^2)$ or the renormalized quantum correction at each order: Whether it can be much smaller compared to $M^2$.
We show that the loop corrections are suppressed if the mass of the field in the corresponding loop is heavy.
\begin{quote}
{\bf Main Theorem} The renormalized self-energy at each order is suppressed as
\be 
\begin{split}
\tilde \Sigma_n^{\text{ren}}(p^2) & \equiv \tilde \Sigma_n(p^2) -\tilde \Sigma_n(m^2) - (p^2-m^2)\frac{d\tilde \Sigma_n}{d (p^2)}(m^2) \\
&= {\cal O} \left(\frac{m^4}{M^2},\frac{p^4}{M^2} \right),
\end{split}
\ee
where $M$ is the mass of the heavy field in the corresponding loop. It vanishes in the decoupling limit $M\to \infty$. 
\end{quote}
It means that the physics of a low-energy scalar field receives suppressed loop corrections and is insensitive to that of UV.
This is an extension of the Appelquist--Carazzone decoupling theorem \cite{Appelquist:1974tg}.

The sketch of the proof goes as follows, with details found in the companion paper \cite{Choi:2024cbs}. The general form of the $n$-loop self-energy amplitude is (we consider one coupling for simplicity, but the multi-coupling generalization is straightforward)
\be 
\tilde \Sigma_n(p^2) = \int d^4k_1 \dots d^4 k_n P(p,k_1,\dots,k_n) \prod_{j=1}^l \frac{1}{L_j(p,k_1,\dots,k_n)-m_j^2} ,
\ee
where $L_j$ are quadratic functions of the momenta and $m_j$ are the mass of the loop propagating fields. $P$ is a polynomial arising from the spins of the fields, whose dependence on $p$ can be moved by redefinition of the internal momenta to a scalar internal line that we always have. Here and in what follows, we suppress the dependence on the coupling $n$ for simplicity \cite{Choi:2024cbs}. 

Feynman parametrization merges the product of the propagators into a single power 
\be 
 [Q(k_1,\dots,k_n)+\Delta(p^2)]^{-l},
\ee 
where $l$ is an integer, $Q(k_1,\dots,k_n)$ is a quadratic form of the internal momenta and $\Delta(p^2)$ is the rest.
By completing the square in all the internal momenta and shifting some momenta, we can always single out the dependence in $p^2$ into the unique combination 
\be \label{Delta}
\begin{split}
 \Delta(p^2) & \equiv - f p^2 + \sum_i g_{i}  m_i^2, \\
 & \equiv - fp^2 + g M^2,
\end{split}
\ee
where $f,g_{i} ,g$ are the coefficients of $p^2, m_i^2,M^2,$ respectively. Every coefficient depends on the Feynman parameter. 
To facilitate the decoupling, we introduced the average mass $M$. If one of $m_i$ is hierarchically large, $M$ is also.

After the momentum integration and cancellation of subdiagram divergence \`a la BPHZ \cite{Bogoliubov:1957gp,Hepp:1966eg,Zimmermann:1967,Lowenstein:1975ps,Lowenstein:1975rg,Weinberg:1959nj} (see also \cite{BS,Blaschke:2013cba,Mooij:2021ojy}), the self-energy has the form
\cite{Choi:2024cbs,Weinberg:1995mt}
\be \label{selfeform}
\tilde \Sigma_n(p^2)= \int d{\bf x} \left[ A + c \Delta(p^2) + e \Delta(p^2) \log \frac{B}{\Delta (p^2)} \right],
\ee
where $c,e$ are order-one constants and ${d \bf x}$ is the measure of the Feynman parameters. We verify it using the fact that $\tilde \Sigma_n(p^2)$ is of mass dimension two and is a function of $\Delta(p^2)$ that is also dimension two.
The dimensionless constant $c$ can appear, depending on the choice of regularization. The quadratic and logarithmic divergences are respectively regularized as $p^2$-independent constants $A$ and $B$, which have mass dimension two. All the constants here depend on the Feynman parameters in general.

The mass correction (\ref{SlidingMass}) that we observe in the experiment is
\be \label{MassCorr} \begin{split}
 \tilde \Sigma_n^{\text{ren}}(p^2)&=  \tilde \Sigma_n(p^2) - \tilde \Sigma_n(m^2) - (p^2-m^2) \frac{d \tilde \Sigma_n}{d p^2}(m^2) \\
 &=    \int d{\bf x}e \left[  \Delta(p^2) \log \frac{\Delta(m^2)}{\Delta(p^2)} - (p^2-m^2)f \right].
 \end{split}
\ee
Not only the regularized divergences $A, B$ but also the scheme-dependent term proportional to $c$ are canceled. Nothing depends on the choice of regularization \cite{Zimmermann:1969jj, Choi:2024cbs}, and we do not need to remove the divergence by hand, like minimal subtraction. The latter is due to that $\Delta$ is linear in $p^2$, and hence the relations 
\be \label{Deltaprel}
 \frac{d \Delta}{d(p^2)}(m^2) =- f, \quad \Delta(p^2) - \Delta(m^2) =- (p^2-m^2)f
\ee
hold.
The quadratic divergence $A$ is canceled in $\tilde \Sigma_n(p^2) - \tilde \Sigma_n(m^2)$ and the logarithmic divergence $B$ is canceled in the first order expansion in $p^2$ \cite{Choi:2023cqs}. 
Since, in the decoupling limit $m^2 \ll M^2$ and $p^2 \ll M^2$,
\be 
\log \Delta(p^2) = \log g M^2 -\frac{f p^2}{g M^2} -\frac{f^2 p^4}{2 g M^4} + {\cal O}\left( \frac{p^6}{M^6} \right),
\ee
the renormalized self-energy reduces as
\be \label{RPM} \begin{split}
 \tilde \Sigma_n^{\text{ren}}(p^2) &=   \int d{\bf x}e \left[ (g M^2-fp^2) \left( \frac{f p^2}{g M^2} +  \frac{f^2 p^4}{ 2 g^2 M^4} + {\cal O}\left( \frac{p^6}{M^6} \right) - \frac{f m^2}{ g M^2} - \frac{f^2 m^4}{2 g^2 M^4} + {\cal O}\left( \frac{m^6}{M^6} \right)   \right) \right.\\ 
 &\qquad \left.- (p^2-m^2)  f \right] \\
 &=  \int d{\bf x} \frac{e f^2}{g} \left[- \frac{p^4}{2 M^2} - \frac{ m^4}{2  M^2}  +{\cal O}\left( \frac{p^6}{M^4} \right) + {\cal O}\left( \frac{m^6}{M^4} \right)   \right] .
\end{split}
\ee
This completes the proof. {\em The cancelation of the quadratic terms in $p$ and $m$ takes place} because of the relation (\ref{Deltaprel}). Note that this removes the naively estimated ${\cal O}(M^2)$ term. 

Therefore, the mass (\ref{SlidingMass}) is perturbatively well-defined. We expect a similar decoupling behavior in the super-renormalizable operators. It exhibits power running, seen from the dependence in $p^2$ \cite{Choi:2023mma}.  
This theorem also holds for vacuum stability, that is, if $m^2<0$ and the mass term serves as a part of the scalar potential. Thus, even if we calculate the scalar mass from vacuum expectation values and their loop corrections at the Grand Unification scale, the scalar mass is stable.

%\subsection*{Acknowledgments}
The author thanks Jong-Hyun Baek, Sungwoo Hong, Hyung-Do Kim, Bumseok Kyae, Hye-Seon Im, Stefan Groot Nibbelink, Hans-Peter Nilles, Ruiwen Ouyang, Mu-In Park, Jaewon Song and Piljin Yi for discussions. 
This work is partly supported by the grant RS-2023-00277184 of the National Research Foundation of Korea.


\begin{thebibliography}{99}

%\cite{Zimmermann:1969jj}
\bibitem{Zimmermann:1969jj}
W.~Zimmermann,
``Convergence of Bogolyubov's method of renormalization in momentum space,''
Commun. Math. Phys. \textbf{15}, 208-234 (1969)
doi:10.1007/BF01645676;

W.~Zimmermann, in 
S.~Deser, M.~Grisaru, H.~Pendleton, Lectures on Elementary Particles and Quantum Field Theory. Volume 1. 1970 Brandeis University Summer Institute in Theoretical Physics, 1970.

%\cite{Choi:2024cbs}
\bibitem{Choi:2024cbs}
K.~S.~Choi,
``Renormalization, Decoupling and the Hierarchy Problem,''
[arXiv:2408.06406 [hep-ph]]. 

%\cite{Gildener:1976ai}
\bibitem{Gildener:1976ai}
E.~Gildener,
``Gauge Symmetry Hierarchies,''
Phys. Rev. D \textbf{14} (1976), 1667
doi:10.1103/PhysRevD.14.1667
%847 citations counted in INSPIRE as of 09 Mar 2024

%\cite{Gildener:1979dd}
\bibitem{Gildener:1979dd}
E.~Gildener,
``GAUGE SYMMETRY HIERARCHIES REVISITED,''
Phys. Lett. B \textbf{92} (1980), 111-114
doi:10.1016/0370-2693(80)90316-0
%72 citations counted in INSPIRE as of 09 Mar 2024

%\cite{Georgi:1993mps}
\bibitem{Georgi:1993mps}
H.~Georgi,
``Effective field theory,''
Ann. Rev. Nucl. Part. Sci. \textbf{43} (1993), 209-252
doi:10.1146/annurev.ns.43.120193.001233
%679 citations counted in INSPIRE as of 28 Oct 2024

%\cite{Georgi:1981vf}
\bibitem{Georgi:1981vf}
H.~Georgi,
``AN ALMOST REALISTIC GAUGE HIERARCHY,''
Phys. Lett. B \textbf{108} (1982), 283-284
doi:10.1016/0370-2693(82)91193-5
%118 citations counted in INSPIRE as of 28 Oct 2024

%\cite{Arkani-Hamed:1998jmv}
\bibitem{Arkani-Hamed:1998jmv}
N.~Arkani-Hamed, S.~Dimopoulos and G.~R.~Dvali,
``The Hierarchy problem and new dimensions at a millimeter,''
Phys. Lett. B \textbf{429} (1998), 263-272
doi:10.1016/S0370-2693(98)00466-3
[arXiv:hep-ph/9803315 [hep-ph]].
%7492 citations counted in INSPIRE as of 28 Oct 2024

%\cite{Weinberg:1978ym}
\bibitem{Weinberg:1978ym}
S.~Weinberg,
``Gauge Hierarchies,''
Phys. Lett. B \textbf{82} (1979), 387-391
doi:10.1016/0370-2693(79)90248-X
%365 citations counted in INSPIRE as of 09 Mar 2024

%\cite{Giudice:2013yca}
\bibitem{Giudice:2013yca}
G.~F.~Giudice,
``Naturalness after LHC8,''
PoS \textbf{EPS-HEP2013} (2013), 163
doi:10.22323/1.180.0163
[arXiv:1307.7879 [hep-ph]].
%153 citations counted in INSPIRE as of 09 Mar 2024

\bibitem{NC}
N.~Craig, ``Naturalness and new approaches to the hierarchy problem,'' (2017). URL: https://www.ias.edu/sites/default/files/pitp/craig.pdf 

%\cite{tHooft:1979rat}
\bibitem{tHooft:1979rat}
G.~'t Hooft,
``Naturalness, chiral symmetry, and spontaneous chiral symmetry breaking,''
NATO Sci. Ser. B \textbf{59} (1980), 135-157
doi:10.1007/978-1-4684-7571-5\_9
%1061 citations counted in INSPIRE as of 31 Jul 2024

\bibitem{Choi:2023cqs}
K.-S.~Choi,
``On the observables of renormalizable interactions,''
J. Korean Phys. Soc. \textbf{84} (2024) no.8, 591-595
doi:10.1007/s40042-024-01025-7
[arXiv:2310.00586 [hep-ph]].

%\cite{Appelquist:1974tg}
\bibitem{Appelquist:1974tg}
T.~Appelquist and J.~Carazzone,
``Infrared Singularities and Massive Fields,''
Phys. Rev. D \textbf{11}, 2856 (1975)
doi:10.1103/PhysRevD.11.2856

\bibitem{Bogoliubov:1957gp}
N.~Bogoliubov and O.~Parasiuk, 
\textit{{
\"Uber die Multiplikation der Kausalfunktionen in der Quantentheorie der Felder}},
Acta Math. \textbf{97}
  (1957) 227--266.

\bibitem{Hepp:1966eg}
K.~Hepp,
 \textit{{Proof of the Bogolyubov-Parasiuk theorem on   renormalization}}, 
Commun. Math. Phys. \textbf{2} (1966) 301--326.

\bibitem{Zimmermann:1967}
W. Zimmermann,
\textit{Local field equation for $A^{4}$-coupling in renormalized perturbation theory},
Commun. Math. Phys. \textbf{6} (1967) 161--188.

%\cite{Lowenstein:1975ps}
\bibitem{Lowenstein:1975ps}
J.~H.~Lowenstein,
``Convergence Theorems for Renormalized Feynman Integrals with Zero-Mass Propagators,''
Commun. Math. Phys. \textbf{47} (1976), 53-68
doi:10.1007/BF01609353

%\cite{Lowenstein:1975rg}
\bibitem{Lowenstein:1975rg}
J.~H.~Lowenstein and W.~Zimmermann,
``The Power Counting Theorem for Feynman Integrals with Massless Propagators,''
Commun. Math. Phys. \textbf{44} (1975), 73-86
doi:10.1007/BF01609059

%\cite{Weinberg:1959nj}
\bibitem{Weinberg:1959nj}
S.~Weinberg,
``High-energy behavior in quantum field theory,''
Phys. Rev. \textbf{118}, 838-849 (1960)
doi:10.1103/PhysRev.118.838;

\bibitem{BS}
Bogoliubov, D. V. Shirkov,  ``Introduction to the Theory of Quantized Fields.''   John Wiley \& Sons Inc; 3rd edition (1980).

%\cite{Blaschke:2013cba}
\bibitem{Blaschke:2013cba}
D.~N.~Blaschke, F.~Gieres, F.~Heindl, M.~Schweda and M.~Wohlgenannt,
``BPHZ renormalization and its application to non-commutative field theory,''
Eur. Phys. J. C \textbf{73} (2013), 2566
doi:10.1140/epjc/s10052-013-2566-8
[arXiv:1307.4650 [hep-th]].

%\cite{Mooij:2021ojy}
\bibitem{Mooij:2021ojy}
S.~Mooij and M.~Shaposhnikov,
``QFT without infinities and hierarchy problem,''
Nucl. Phys. B \textbf{990} (2023), 116172
doi:10.1016/j.nuclphysb.2023.116172
[arXiv:2110.05175 [hep-th]].



%\cite{Weinberg:1995mt}
\bibitem{Weinberg:1995mt}
S.~Weinberg,
``The Quantum theory of fields. Vol. 1: Foundations,''
Cambridge University Press, 2005,
ISBN 978-0-521-67053-1, 978-0-511-25204-4
doi:10.1017/CBO9781139644167

%\cite{Choi:2023mma}
\bibitem{Choi:2023mma}
K.~S.~Choi,
``Exact renormalization of the Higgs field,''
Phys. Rev. D \textbf{109} (2024) no.7, 076008
doi:10.1103/PhysRevD.109.076008
[arXiv:2310.10004 [hep-th]].

\end{thebibliography}
\end{document}